\documentclass[%
aip,jap,
 amsmath,amssymb,floatfix,
 preprint,%
]{revtex4-1}
\usepackage{multirow}
\usepackage{amsmath,amssymb,amsfonts}
\usepackage{algorithmic}
\usepackage{array}
\usepackage{xcolor}
\usepackage{footnote}
\usepackage{threeparttable}
\usepackage{graphicx}
\usepackage{textcomp}
\usepackage{xcolor}
\usepackage{array}
\usepackage{mathtools}
\usepackage{siunitx}

\DeclarePairedDelimiter\abs{\lvert}{\rvert}%

\def\BibTeX{{\rm B\kern-.05em{\sc i\kern-.025em b}\kern-.08em
    T\kern-.1667em\lower.7ex\hbox{E}\kern-.125emX}}
\begin{document}

\title{Spin Wave Based $4$-$2$ Compressor
{}
}

\author{Abdulqader Mahmoud}
\email{A.N.N.Mahmoud@tudelft.nl}
\affiliation{Delft University of Technology, Department of Quantum and Computer Engineering, 2628 CD Delft, The Netherlands}

\author{Frederic Vanderveken}
\affiliation{KU Leuven, Department of Materials, SIEM, 3001 Leuven, Belgium}
\affiliation{Imec, 3001 Leuven, Belgium}

\author{Florin Ciubotaru}
\affiliation{Imec, 3001 Leuven, Belgium}

\author{Christoph Adelmann}
\affiliation{Imec, 3001 Leuven, Belgium}

\author{Sorin Cotofana}
\affiliation{Delft University of Technology, Department of Quantum and Computer Engineering, 2628 CD Delft, The Netherlands}

\author{Said Hamdioui}
\email{S.Hamdioui@tudelft.nl}
\affiliation{Delft University of Technology, Department of Quantum and Computer Engineering, 2628 CD Delft, The Netherlands}

\begin{abstract}
By their very nature, Spin Waves (SWs) consume ultra-low amounts of energy, which makes them suitable for ultra-low energy consumption applications. In addition, a compressor can be utilized to further reduce the energy consumption and enhance the speed of a multiplier. Therefore, we propose a novel energy efficient SW based $4$-$2$ compressor consisting of $4$ XOR gates and $2$ Majority gates. The proposed compressor is validated by means of micromagnetic simulations and compared with the state-of-the-art SW, \SI{22}{nm} CMOS, Magnetic Tunnel Junction (MTJ), Domain Wall Motion (DWM), and Spin-CMOS technologies. The performance evaluation shows that the proposed compressor consumes $2.5$x less and $1.25\times$ less energy than the \SI{22}{nm} CMOS and the conventional SW compressor, respectively, whereas it consumes at least $3$ orders of magnitude less energy than the MTJ, DWM, and Spin-CMOS designs. Furthermore, the compressor achieves the smallest chip real-estate. In summary, the performance evaluation of our proposed compressor shows that the SW technology has the potential to progress the state-of-the-art circuit design in terms of energy consumption and scalability. 
\end{abstract}

\maketitle

\section{Introduction}

Complementary Metal Oxide Semiconductor (CMOS) downscaling has been efficient to meet the exploding market requirements for highly efficient computing platforms that process the raw data resulting from the information technology revolution \cite{data1}. However, CMOS downscaling becomes very difficult as we approach the end of Moore's law because of the leakage, cost, and reliability walls \cite{cmosscaling1}. Therefore, researchers have explored different technologies including spintronics \cite{ITRS}. One of the spintronic promising technologies is the Spin Wave (SW) technology because it has ultra-low energy consumption, acceptable delay, and high scalablility\cite{amahmoud2,amahmoud1,parallelism,fanout10,wavepipeline}. As a result, there is a strong interest in  designing SW based circuits.  

Researchers have designed different logic gates and circuits using SWs  \cite{logic21,fanout, parallelism,parallelism1, fanout10, fanout11, logic1,amahmoud1,memory3}. A Mach-Zehnder interferometer was utilized to build the first experimental SW NOT gate \cite{logic21}. Afterwards, single output Majority, (N)AND, (N)OR, and X(N)OR gates were built using Mach-Zehnder interferometers \cite{logic21}, whereas multi-output logic gates were suggested \cite{fanout, fanout10,fanout11}. Moreover, multi-frequency logic gates were reported \cite{parallelism,parallelism1}. On a bigger scale, multiple circuits have been introduced at the conceptual  \cite{logic1}, simulational \cite{amahmoud1}, and also practical millimeter scale level \cite{memory3}. To conclude, SW circuit design is still in its genesis stage. Therefore, the design, validation and demonstration of SW based circuits at different complexity scales is of great interest to progress SW computing.

Driven by the aforementioned information, we propose, validate, and assess a novel SW based $4$-$2$ compressor consisting of $4$ XOR and $2$ Majority gates. In the following, we summarize the main contributions of the paper: 
\begin{itemize}
  \item Designing a novel SW $4$-$2$ compressor.
  \item Validating the proposed $4$-$2$ Compressors by means of micromagnetic simulations.  
  \item Demonstrating the compressor superiority by comparing its performance with the state-of-the-art SW, \SI{22}{nm} CMOS, Magnetic Tunnel Junction (MTJ), Domain Wall Motion (DWM), and Spin-CMOS technologies. The evaluation results show that the proposed compressor consumes consumes $1.25$x less energy than the conventional SW compressor, and $2.5$x less energy than the \SI{22}{nm} CMOS counterparts. In addition, it outperforms the MTJ, DWM, and Spin-CMOS designs by at least $3$ orders of magnitude. Furthermore, it achieves the smallest chip real-estate. 
\end{itemize}

The paper is organized as follows. We explain the SW background and computing paradigm in Section \ref{sec:Basics of spin-wave technology}. Next, we illustrate the proposed compressor in Section \ref{sec:Proposed approximate functions}, and present the simulation setup, results, and performance evaluation in Section \ref{sec:Simulation Setup and Results}. Section \ref{sec:Conclusion} concludes the paper.

\section{Spin Wave Based Technology Fundamental and Computing Paradigm}
\label{sec:Basics of spin-wave technology}

The magnetization dynamics in a ferro- or ferrimagnetic material is described by the Landau-Lifshitz-Gilbert (LLG) equation  \cite{amahmoud2}: $\frac{d\vec{M}}{dt} =-\abs{\gamma} \mu_0 \left (\vec{M} \times \vec{H}_{eff} \right ) + \frac{\alpha}{M_s} \left (\vec{M} \times \frac{d\vec{M}}{dt}\right )$, where $\gamma$ is the gyromagnetic ratio, $\mu_0$ the vacuum permeability, $M$ the magnetization, $M_s$ the saturation magnetization, $\alpha$ the damping factor, and $H_{eff}$ the effective field consisting of the external field, the exchange field, the demagnetizing field, and the magneto-crystalline field.

For small magnetic disturbances, the LLG equation predicts wave-like magnetic motion. These wave-like solutions are called Spin Waves (SWs), reflecting collective excitations of the magnetization within the magnetic material \cite{amahmoud2}. 

The SW amplitude and phase can be used to encode information at different frequencies \cite{amahmoud2,parallelism}. Moreover, the processing of this information is performed by the interference principle. For example, if two SWs with the same amplitude, wavelength, and frequency meet in the waveguide, they interfere constructively if they have the same phase, \emph{i.e.} $\Delta \phi=0$, and destructively if they have opposite phases, \emph{i.e.} $\Delta \phi=\pi$. In addition, SWs naturally support Majority gates because the interference of an odd number of SWs is based on the Majority decision. For instance, if $3$ SWs with the the same amplitude, wavelength, and frequency meet in the same waveguide, the interference result is a SW with phase $0$ if at least $2$ SWs have a phase of $0$, whereas the interference result is a SW with phase $\pi$ if at least $2$ SWs have a phase of $\pi$. Note that such an implementation in CMOS technology requires $18$ transistors whereas it can be directly implemented in SW technology \cite{amahmoud2}. In this paper, logic $0$ corresponds to a SW with phase $0$, whereas logic $1$ corresponds to a SW with phase $\pi$. 

SW device consists of four main stages: i) excitation stage, ii) waveguide, iii) functional stage, and iv) detection stage \cite{amahmoud2}. At the excitation stage, SW is excited by means of voltage driven cells such as Magneto-Electric (ME) cells or current driven cells such as inductive antennas \cite{amahmoud2}. After SW excitation, SW propagates through the waveguide which can be made of different material such as Permalloy, and CoFeB. At the functional stage, the SW can be manipulated, normalized, amplified, or interfered with other SWs. Finally, the resultant SW is captured at the detection stage, which can be similar or different than the ones utilized at the excitation stage. Two different techniques can be utilized to capture the SW: phase detection and threshold detection. In the phase detection, the resultant SW phase is compared with a predefined phase, if its phase is $0$, the output is logic $0$, and otherwise, logic $1$. On the other hand, in the threshold detection, the dynamic magnetization amplitude is compared with a predefined threshold, i.e., if the amplitude is larger than the predefined threshold, the output is logic $0$, and otherwise, logic $1$ \cite{amahmoud2}.

\begin{figure}[t]

\centering
  \includegraphics[width=0.6\linewidth]{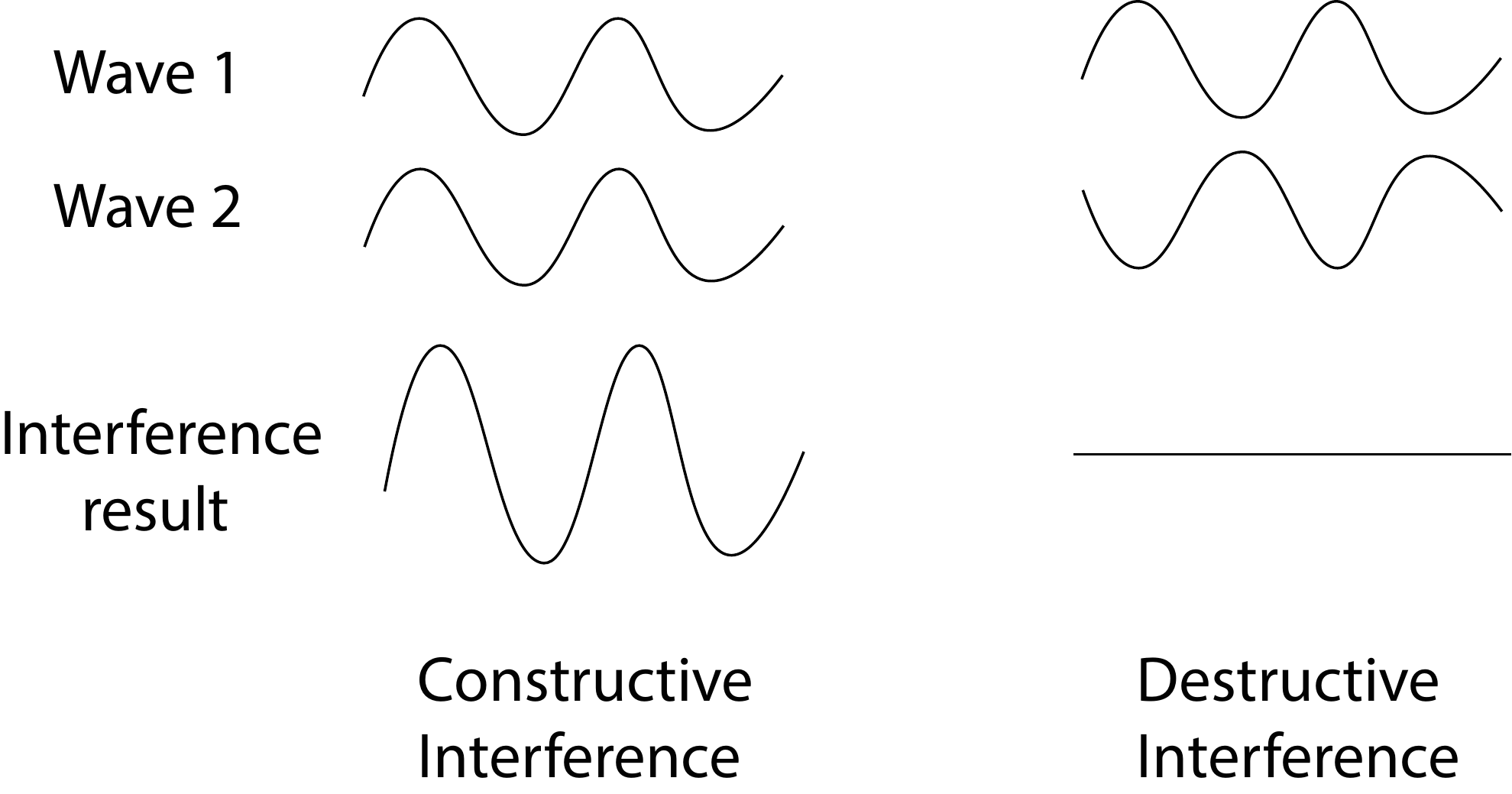}
 
  \caption{SW Interference.}

  \label{fig:interference}
 \end{figure}

\section{SW $4$-$2$ Compressor}
\label{sec:Proposed approximate functions}
The fast multiplier consists of three main stages: partial product generator, partial products reducer, and final production computer; the main part of the energy consumption and delay originates from the partial product stage. This can be optimized by utilizing a $4$-$2$ compressor in the partial products reducer stage of the fast multiplier \cite{conventional_compressor2}. Therefore, we built a SW $4$-$2$ compressor.

\begin{figure}[t]

\centering
  \includegraphics[width=\linewidth]{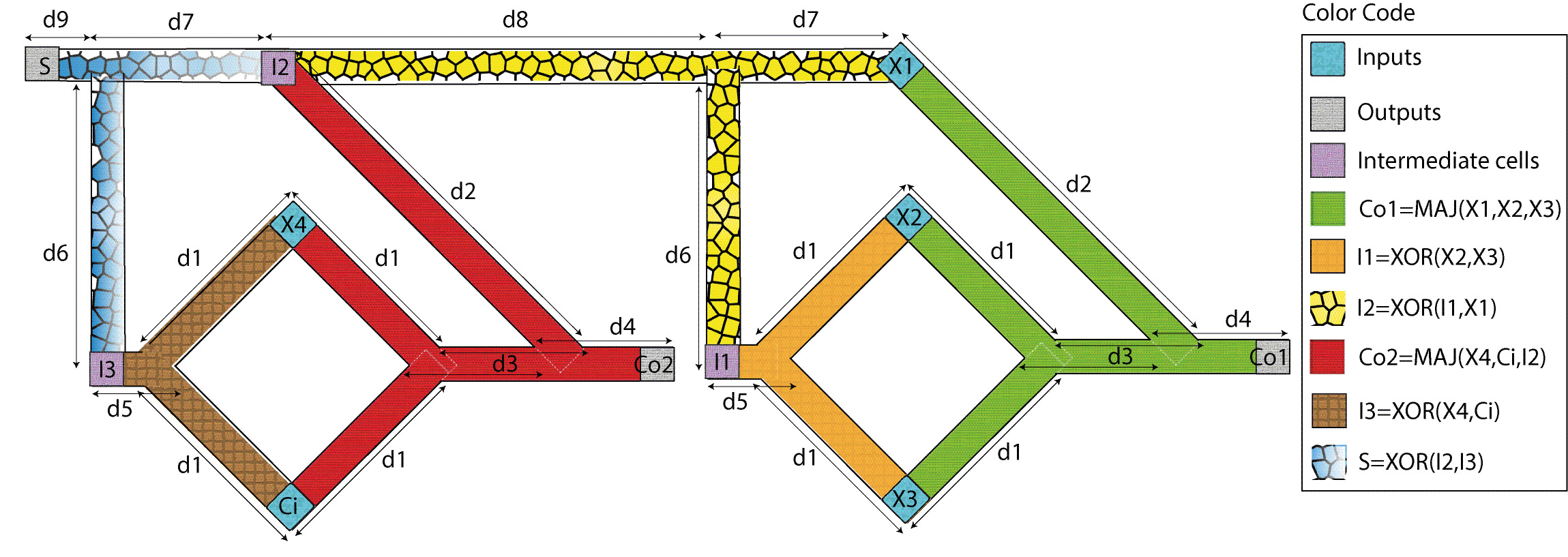}

  \caption{Spin Wave Based $4$-$2$ Compressor.}

  \label{fig:structure1}
\end{figure}

Figure \ref{fig:structure1} presents the proposed $4$-$2$ compressor consisting of $5$ inputs $X1$, $X2$, $X3$, $X4$, and $C_i$ and $3$ outputs $C_{o1}=MAJ(X1,X2,X3)$, $C_{o2}=MAJ(XOR(XOR(X1,X2),X3),X4,C_i)$, and $S=XOR(XOR(XOR(XOR(X1,X2),X3),X4),C_i)$ in addition to $3$ intermediate cells $I_1$, $I_2$, and $I_3$, which are repeaters to receive and excite the SWs with the suitable amplitude and phase.

In order to ensure the correct functionality of the proposed $4$-$2$ compressor, all SWs must be excited at the same amplitude, wavelength, and frequency. The SW wavelength must be larger than the waveguide width to simplify the interference pattern. Moreover, the structure must be designed carefully to guarantee the correct functionality of the compressor because the structure's dimension affects the interference results. For example, if constructive interference is required at the intersection point when the waves have the same phase and destructive interference otherwise, then the device dimensions $d_1$,$d_2$,$d_3$,$d_5$,$d_6$,$d_7$,and $d_8$ must equal to $n\lambda$ where $n=0,1,2,\ldots$ Note that this is the case in our design. The outputs $C_{o1}$ and $C_{o2}$ must be located at a specific position as they are based on phase detection. Hence, by changing its location, it is feasible to extract the inverted output or the non-inverted output. For example, if the desired result is to capture the non-inverted output, the distance $d_4$ must equal $n\lambda$ which is the case for $C_{o1}$ and $C_{o2}$. On the other hand, as the output $S$ is detected based on threshold detection, the resultant SW is compared with a predefined threshold value as previously discussed. To detect the largest possible SW amplitude, the output $S$ must be located as close as possible to the interference point, i.e., $d_9$ must be as small as possible.

The proposed $4$-$2$ SW compressor works as follows: \begin{itemize}
\item Carry-out1 output $C_{o1}$: The SWs excited at $X2$ and $X3$ interfere constructively or destructively depending on their phase at the intersection point. Then the SW interference result propagates further through the waveguide and interferes with the SW excited at $X1$ at the intersection point between the waveguides. Finally, the resultant SW is captured at $C_{o1}$ based on phase detection. 
\item Carry-out2 output $C_{o2}$: The SWs excited at $X2$ and $X3$ interfere constructively or destructively depending on their phase at the intersection point. After that, the resultant wave is received by repeater $I1$ which will excite a SW with a suitable phase depending on the received SW magnetization. If the received SW magnetization is larger than a threshold, a SW with phase of $0$ will be excited, and a SW with phase of $\pi$ will be excited, otherwise. Then, the SW excited from $I1$ interferes with the SW excited from $X3$. Next, the resultant SW will be received by the repeater $I2$ which will excite a SW with a suitable phase depending on the received SW magnetization at the intersection point between the waveguides. Meanwhile, the SWs excited from $X4$ and $Ci$ will interfere at the intersection point. Finally, the resultant SW will interfere with the SW excited from $I2$, and the result will be captured by $C_{o2}$ based on phase detection. 
\item Sum output $S$: The SWs excited from $X4$ and $Ci$ will interfere at the intersection point between the two waveguides, and the result will be detected by repeater $I3$. Next,  repeater $I3$ will excite a SW with a suitable phase depending on the received SW magnetization as previously discussed. Finally, the output $S$ will capture the results of the interference between SWs excited from $I2$ and $I3$ based on threshold detection.
\end{itemize}

\section{Simulation Setup, Results and Performance Evaluation}
\label{sec:Simulation Setup and Results}

\subsection{Simulation Setup}
We utilized the following parameters to validate the proposed structure by MuMax3 \cite{mumax}: \SI{50}{nm} wide and \SI{1}{nm} thick $Fe_{60}Co_{20}B_{20}$ waveguide with saturation magnetization $M_s$ of \SI{1.1}{MA/m}, damping constant $\alpha$ of $0.004$, and exchange stiffness $A_{exch}$ of \SI{18.5}{pJ/m} \cite{parameters}. We excited the SWs with a \SI{10}{GHz} Gaussian pulse with sigma of \SI{500}{ps} to save energy, guarantee a single frequency SW excitation, and achieve high group velocity. The wavenumber $k$ is determined from the SW dispersion relation, which makes the wavelength equals to $\lambda$=$2\pi/k$=\SI{170}{nm}. As discussed in Section \ref{sec:Proposed approximate functions}, the distances $d_1$, $d_2$, $d_3$, $d_6$, $d_7$, and $d_8$ equal to $n\lambda$. The distances were determined to be: $d_1$=\SI{340}{nm}(n=2), $d_2$=\SI{850}{nm}(n=5), $d_3$=\SI{680}{nm}(n=4), $d_4$=\SI{170}{nm}(n=1), $d_5$=\SI{50}{nm}, $d_6$=\SI{340}{nm}(n=2), $d_7$=\SI{340}{nm}(n=2), $d_8$=\SI{1020}{nm}(n=6) and $d_9$=\SI{50}{nm}.

\subsection{Simulation Results}
\begin{figure}[t]
\centering
  \includegraphics[width=\linewidth]{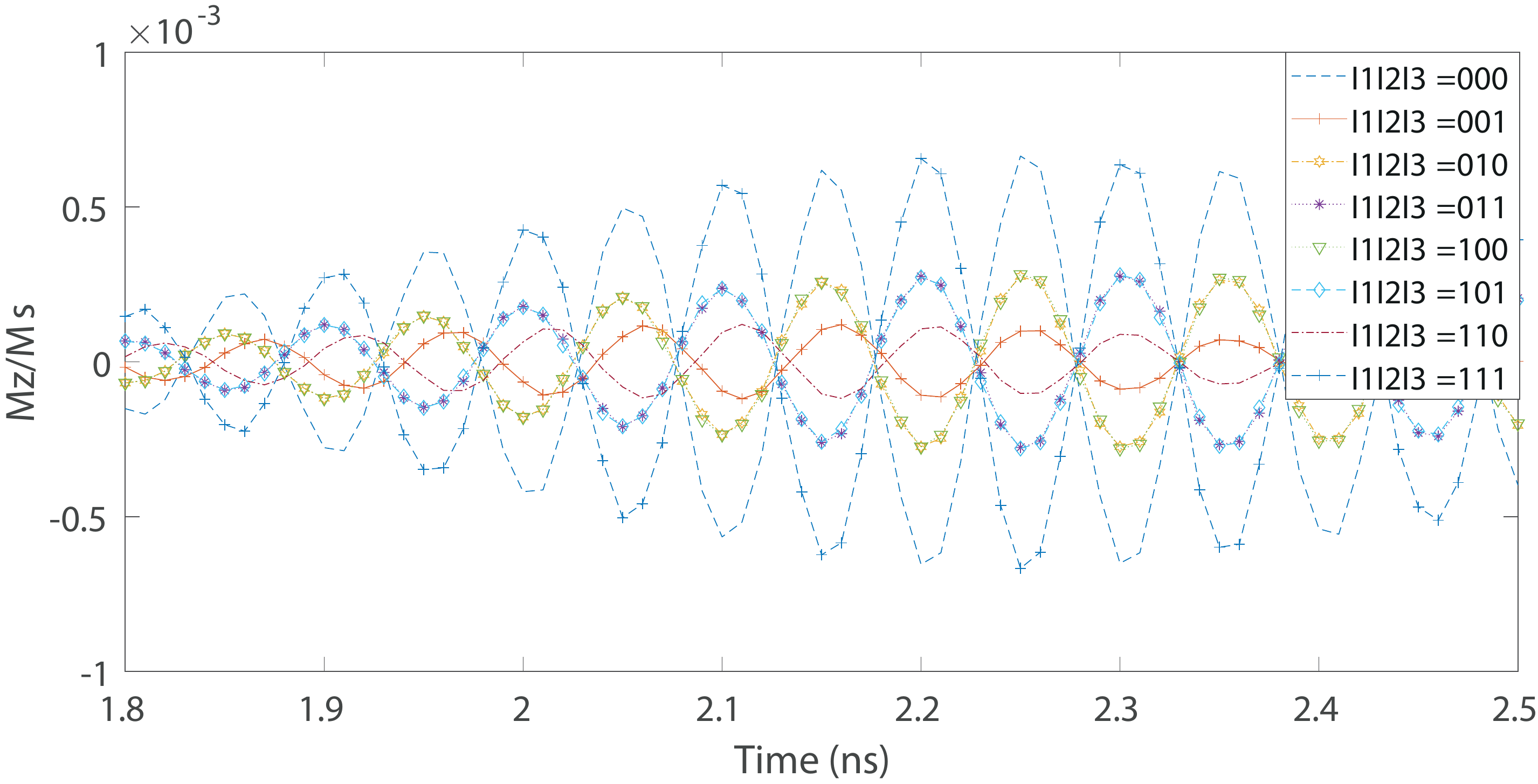}

  \caption{Normalized $4$-$2$ Compressor Carry-out1 Output $C_{o1}$.}
  \label{fig:result}

\end{figure}

Figure \ref{fig:result} presents the proposed compressor carry-out1 $C_{o1}$ MuMax3 simulation results for \{$X_1$,$X_2$,$X_3$\}= \{$0$,$0$,$0$\}, \{$0$,$0$,$0$\}, \{$0$,$0$,$1$\}, \{$0$,$1$,$0$\}, \{$0$,$1$,$1$\}, \{$1$,$0$,$0$\}, \{$1$,$0$,$1$\}, \{$1$,$1$,$0$\}, and \{$1$,$1$,$1$\}, respectively. Inspecting the figure, the $C_{o1}$ is captured correctly based on phase detection. For example, $C_{o1}=0$ for  \{$I_1$,$I_2$,$I_3$\}= \{$0$,$0$,$0$\}, \{$0$,$0$,$1$\}, \{$0$,$1$,$0$\}, and \{$1$,$0$,$0$\}, whereas $C_{o1}=1$ for  \{$X_1$,$X_2$,$X_3$\}= \{$0$,$1$,$1$\}, \{$1$,$0$,$1$\}, \{$1$,$1$,$0$\}, and \{$1$,$1$,$1$\} at time=\SI{2.25}{ns}.

Table \ref{table:3} presents the normalized magnetization of the SW received by the repeater $I_1$ and the SW excited by $I_1$ for different input combinations \{$X_2$,$X_3$\}= \{$0$,$0$\}, \{$0$,$1$\}, \{$1$,$0$\}, and \{$1$,$1$\}, respectively. Note that the threshold technique is used to detect and excite the SW at $I_1$ such that if the SW magnetization is larger than the threshold, $I_1$ excites a SW with $\phi=0$, whereas otherwise, $I_1$ excites a SW with $\phi=\pi$. The threshold is calculated by averaging the two nearest cases, \emph{i.e.} \{$X_2$,$X_3$\}= \{$1$,$0$\}, \{$1$,$1$\}, resulting in $0.585$ for this case. Inspecting the table, we can see that the SW magnetization received by $I_1$ is larger than $0.585$ for the input combinations \{$X_2$,$X_3$\}= \{$0$,$0$\}, and \{$1$,$1$\}, whereas the SW magnetization received by $I_1$ is less than $0.585$ for the input combinations \{$X_2$,$X_3$\}= \{$0$,$1$\}, and \{$1$,$0$\}. 

The same reasoning holds for $I_2$ for which the results are presented in Table \ref{table:4}. Here, the threshold is set to $0.82$ which is the average of the two cases \{$X_2$,$X_3$\}= \{$0$,$1$\}, \{$1$,$1$\}. Inspecting the table, we can see that the SW magnetization received by $I_2$ is larger than $0.82$ for the input combinations \{$X_1$,$I_1$\}= \{$0$,$0$\}, and \{$1$,$1$\}, whereas the SW magnetization received by $I_2$ is less than $0.82$ for the input combinations \{$X_1$,$I_1$\}= \{$0$,$1$\}, and \{$1$,$0$\}. After that, the same results are obtained for $C_{o2}$ which is detected based on phase detection as $C_{o1}$, and $I_3$ and $S$ which are detected based on threshold detection as $I_{1}$ and $I_2$ with the same analysis.

Therefore, the micromagnetic simulation results demonstrated that the $4$-$2$ SW compressor is functioning correctly.

\begin{table}[t]
\caption{Normalized SW Magnetization at $I1$}

\label{table:3}
\centering
  \begin{tabular}{|>{\centering}m{5em}|>{\centering}m{13em}|>{\centering}m{8em}|}
    \hline
    Inputs ($X2X3$) & Normalized SW Magnetization received by $I1$  &  SW excited by $I1$ \tabularnewline
    \hline
    $00$ & $1$  & SW with $\phi=0$ \tabularnewline \hline
    $01$ & $0.18$ & SW with $\phi=\pi$ \tabularnewline \hline
    $10$ & $0.18$ & SW with $\phi=\pi$  \tabularnewline \hline
    $11$ & $0.99$ & SW with $\phi=0$  \tabularnewline \hline
  \end{tabular}

\end{table}

\begin{table}[t]
\caption{Normalized SW Magnetization at $I2$}

\label{table:4}
\centering
  \begin{tabular}{|>{\centering}m{5em}|>{\centering}m{13em}|>{\centering}m{8em}|}
    \hline
    Inputs ($X1I1$) & Normalized SW Magnetization received by $I2$  &  SW excited by $I2$ \tabularnewline
    \hline
    $00$ & $1$  & SW with $\phi=0$ \tabularnewline \hline
    $01$ & $0.65$ & SW with $\phi=\pi$ \tabularnewline \hline
    $10$ & $0.64$ & SW with $\phi=\pi$  \tabularnewline \hline
    $11$ & $0.99$ & SW with $\phi=0$  \tabularnewline \hline
  \end{tabular}

\end{table}

\subsection*{Performance Evaluation}
In order to assess the performance of the proposed $4$-$2$ SW compressor and see the potential of such an approach, we evaluate it and compare it with the state-of-the-art SW, \SI{22}{nm} CMOS \cite{CompCMOS1}, Magnetic Tunnel Junction (MTJ) \cite{SPIN}, Domain Wall Motion (DWM) \cite{SPIN}, and Spin-CMOS \cite{SPIN} technologies in terms of energy, delay, and area. We have made the following assumptions for the performance evaluation \cite{amahmoud1}: (i) The excitation, detection, and repeater cells are Magnetoelectric (ME) cells, and their power consumption, and delay are \SI{34}{nW}, and \SI{0.42}{ns}, respectively. (ii) SWs do not consume noticeable energy while interferring with each other or propagating in the waveguide. Note that these assumptions might need re-evaluation in the near future as SW technology is still in its infancy stage. 

Table \ref{table:5} presents the performance evaluation of the proposed compressor, and the comparison with the state-of-the-art. As it can be observed from the table, the proposed SW compressor consumes $2.5$x less energy than the \SI{22}{nm} CMOS counterpart while requiring $119$x more delay \cite{CompCMOS1}. In addition, the proposed SW compressor consumes at least $3$ orders of magnitude less energy than the MTJ, DWM, and Spin-CMOS counterparts, while requiring $1.84$x, and $1.26$ more delay, and $1.28$x less delay than the MTJ, DWM, and Spin-CMOS counterparts, respectively \cite{SPIN}. When compared with the conventional SW $4$-$2$ compressor, which is two cascaded full adders, the proposed SW $4$-$2$ compressor consumes $1.25$x less energy than the conventional SW $4$-$2$ compressor while needing $1.22$x less delay. Moreover, the proposed compressor requires the least number of devices in comparison with the other designs as can be seen in Table \ref{table:5}. Note that the SW delay can be improved by using other materials which have higher group velocity.

\begin{table}[t]
\caption{$4$-$2$ Compressor Performance Comparison}

\label{table:5}
\centering
  \begin{tabular}{|c|c|c|c|c|}
    \hline
    Design&Technology & Energy (fJ)  &  Delay (ns) & Device No. \tabularnewline
    \hline
    \cite{CompCMOS1} & CMOS  & $0.4$ & $0.048$ & $38$ \tabularnewline \hline
    \cite{SPIN}& MTJ  & $85680$  & $20.4$ & $76$ \tabularnewline \hline
    \cite{SPIN}& DWM& $630$  & $3.7$ & $58$ \tabularnewline \hline
    \cite{SPIN}& Spin-CMOS & $667$  & $6$ & $68$ \tabularnewline \hline
    Conventional SW & Spin Wave & $0.2$  & $5.72$ & $14$ \tabularnewline \hline
    Proposed SW & Spin Wave & $0.16$  & $4.68 $ & $11$ \tabularnewline \hline
  \end{tabular}

\end{table}

We assessed the proposed SW compressor on an application level utilizing the JPEG compression algorithm to see the potential of such an approach in larger scale. In the JPEG algorithm \cite{JPEG}, DCT and IDCT can be implemented using the $4$-$2$ compressor \cite{SPIN}. If we implemented the DCT and IDCT by means of the proposed $4$-$2$ SW compressor, we expect to achieve ultra-low-energy consumption. As we discussed previously, the proposed compressor consumes $3$ magnitude orders less energy than the Spin-CMOS counterpart which indicates that the DCT/IDCT based on the proposed $4$-$2$ SW compressor will consume at least $3$ orders of magnitude less energy than the DCT/IDCT based on the Spin-CMOS $4$-$2$ SW compressor \cite{SPIN}.

In this paper, our main goal is to propose and validate the SW compressor as a proof of concept without considering thermal noise and variability effects. However, in \cite{DC}, it was presented that the thermal noise, the edge roughness and the waveguide trapezoidal cross section do not have noticeable effects on the gate's functionality. Therefore, we expect that the thermal noise and variability will have limited effect on the compressor. Nevertheless, we will investigate such phenomena in the future. 

It was shown that SW technology can be very effective and has the requirements to progress the state-of-the-art in terms of energy consumption and scalability. However, some open issues are still to be solved \cite{amahmoud2}. For example, although Magneto-Electric (ME) cells seem to be the right choice for the SW excitation and detection, their efficient behavior is not yet been  experimentally realized. Moreover, although SW technology is highly scalable as the only limitation for a SW device scalability is the SW wavelength, the SW has not yet been distinguished from the noise at the nano-scale \cite{amahmoud2}. However, we are sure that the industry, as always, will find its way to efficient nanoscale SW devices and benefit from the SW computing paradigm.

\section{Conclusions}
\label{sec:Conclusion}

We proposed and validated by means of micro-magnetic simulation a novel $4$-$2$ Spin Wave (SW) compressor. The proposed compressor was assessed and compared with the state-of-the-art SW, \SI{22}{nm} CMOS, Magnetic Tunnel Junction (MTJ), Domain Wall Motion (DWM), and Spin-CMOS technologies. The evaluation result showed that the proposed compressor consumed $2.5$x less energy than \SI{22}{nm} CMOS counterpart. In addition, it outperformed the MTJ, DWM, and Spin-CMOS designs by at least $3$ orders of magnitude. Moreover, it consumed $1.25$x less energy than the conventional SW compressor. Furthermore, it achieved the smallest chip real-estate. To conclude, the performance evaluation shows that SW technology has the potential to further progress the circuit design in terms of energy and scalability.

\section*{Acknowledgement}
This work has received funding from the European Union's Horizon 2020 research and innovation program within the FET-OPEN project CHIRON under grant agreement No. 801055.

\bibliography{references}

\end{document}